\documentclass{aa}
\usepackage{graphics}
\begin{document}

\thesaurus{06     
              (08.04.1;  
               08.06.2;  
               08.16.5;  
               08.22.4;  
	      )} 
\title{Revisiting Hipparcos data for pre-main sequence stars
\thanks{Based on observations made with the
ESA Hipparcos astrometry satellite}}
\author{Claude Bertout \inst{1} \and
No\"el Robichon \inst{2,3} \and Fr\'ed\'eric Arenou \inst{2} }

\offprints{Claude Bertout}

\institute{Institut d'Astrophysique de Paris 98bis Bd Arago, 75014
Paris, France; email: bertout@iap.fr
\and
DASGAL, Observatoire de Paris /
CNRS UMR 8633, F-92195 Meudon CEDEX, France,
email: Noel.Robichon@obspm.fr, Frederic.Arenou@obspm.fr
\and
Sterrewacht Leiden, Postbus 9513, NL-2300 RA Leiden, The Netherlands}

\date{Received ; accepted }

\maketitle

\begin{abstract}
We cross-correlate the Herbig \& Bell and Hipparcos Catalogues in order to
extract the results for young stellar objects (YSOs). We compare the
distances of individual young stars and the distance of their presumably
associated molecular clouds, taking into account post-Hip\-parcos distances to
the relevant associations and using Hipparcos intermediate astrometric data
to derive new parallaxes of the pre-main sequence stars based on their
grouping. We confirm that YSOs are located in their associated clouds, as
anticipated by a large body of work, and discuss reasons which make the
individual parallaxes of some YSOs doubtful. We find in particular that the
distance of Taurus YSOs \textit{as a group} is entirely consistent with the
molecular cloud distance, although Hipparcos distances of some faint
Taurus-Auriga stars must be viewed with caution. We then improve some of the
solutions for the binary and multiple pre-main sequence stars. In
particular, we confirm three new astrometric young binaries discovered by
Hipparcos: \object{RY Tau}, \object{UX Ori}, and \object{IX Oph}.

\keywords {stars:pre-main sequence; stars:
formation; stars: distances; stars: variables: general}
\end{abstract}

%

\section{Introduction}


Star formation is widely thought to occur in molecular clouds, which then
provide the raw material from which stars can accumulate during the course
of the gravitational collapse. While many details of the star formation
process remain topics of controversy, the idea that most young stellar
objects (YSOs) should be physically associated with molecular clouds
remained unchallenged until recently.

Two observational findings in the last few years did, however, cast some
doubt on this widely held belief. First came the discovery by the ROSAT
satellite of a population of X-ray active pre-main sequence stars -- many
identified as weak-emission line T Tauri stars (WTTSs) -- extending far
beyond the boundaries of known star-forming regions (e.g., Neuh\"{a}user
et al. \cite{Neuhauser et al. 1995a}, Neuh\"{a}user et al.
\cite{Neuhauser et al. 1995b}). This result was at first
taken as an indication that YSOs could migrate far away from the region of
their formation on short time-scales. Propositions to explain this fact
included the formation of stars in fast-moving molecular cloudlets
(Feigelson \cite
{Feigelson 1996}) or tidally-induced escapes within multiple systems
(Armitage \& Clarke \cite
{Armitage & Clarke 1997}). However, the recent finding that many of these
pre-main sequence objects are most likely the low-mass counterpart of the
Gould Belt OB\ associations
(Guillout et al. \cite{Guillout et al. 1998}) reconciles the
ROSAT results with the conventional idea that these stars formed in (now
dispersed) molecular clouds.

Another observation which may contradict the hypothesis that YSOs and
molecular clouds are associated was recently reported by
Favata et al. (\cite{Favata et al.
1998}). On the basis of the Hipparcos satellite astrometric measurements,
these authors assert that some classical T Tauri stars (CTTSs) of the
Taurus-Auriga star-forming region are apparently much closer than previously
thought and may not belong to the Taurus clouds. If this result were proven
true beyond reasonable doubt, consequences for the physics of star formation
and evolution of solar-type stars would be far-reaching, since arguments for
the youth of a given stellar object relate primarily to its location in the
vicinity of molecular clouds.

Four arguments are used to infer pre-main sequence status of a solar-type
star:
\begin{itemize}
\item  association with OB stars (e.g., in the Orion Trapezium region);
\item  association with dark or bright nebulosity (e.g., in the
Taurus-Auriga region);
\item  location above the main-sequence in the Hertzsprung-Russell Diagram;
\item  presence in the spectrum of the $\lambda 6707$\AA\ LiI resonance line
(with equivalent width larger than in ZAMS stars of the same spectral type).
\end{itemize}

The first two criteria are straightforward: OB associations are short-lived,
which guarantees the youth of associated low-mass objects, while physical
association of a star with a cloud is either seen at the telescope when
close reflection and emission nebulae are present -- as is the case in the
vicinity of many CTTSs -- or inferred from kinematic studies (e.g.
Jones \& Herbig \cite
{Jones & Herbig 1979}).

The last two criteria above are more indirect. The location in the H-R
diagram depends on the assumption made about the respective luminosities of
photosphere and circumstellar matter in a given object (Kenyon \&
Hartmann \cite{Kenyon & Hartmann 1990}), as well as on the assumed distance.
Lithium is, in theory,
destroyed early in the star history, when the temperature at the bottom of
the convection zone reaches about $2\times 10^{6}$K, but the lithium
abundance may be dependent on variables other than age (cf., e.g.,
Ventura et al. 1998).

If some CTTSs appear to be located nearby, and thus far away from known
star-forming region, as suggested by Favata et al. (1998), one must
then choose between two equally unsettling possibilities: (a) either they
are young stars, and one must explain how they arrived at their present
location, or (b) they are field stars that mimic YSOs, and one must
understand how this is possible. Needless to say, there are no obvious
answers to these questions, and the entire picture of low-mass star
formation would have to be fully revised in order to account for these
observations.

This paper provides a detailed re-examination of Hipparcos satellite data
for TTSs, focusing mainly on distances and binarity/multiplicity properties.
Data are discussed in Sect. 2. Sect. 3 then examines the Hipparcos results
relevant to YSO and molecular cloud distances, and concludes that Hipparcos
distances for young stars are generally consistent with their expected
physical association with molecular clouds. The same conclusion was reached
independently by Wichmann et al.
(\cite{Wichmann et al. 1998}), but we extend their analysis
by providing new astrometric solutions for groups of TTSs in various
star-forming regions. We then study the binarity/multiplicity Hipparcos data
in Sect. 4. There, we discuss or improve the astrometric data reductions
for several YSO systems, using Hipparcos intermediate astrometric data and,
when available, taking advantage of recent spectroscopic information.


\section{Hipparcos Catalogue data for pre-main sequence stars}


The sample of young stars contained in the Herbig \& Bell (1988)
Catalogue of Orion population objects with emission lines (HBC) and observed
by the Hipparcos satellite is fairly limited:13\ Herbig Ae/Be stars (7 of
which have significant parallax values); 16 CTTSs, 10 of which have
significant parallaxes; 7 WTTSs or SU Aurigae stars, 4 of them are
positively detected; and 9 stars with uncertain pre-main sequence status, 4
of which have significant parallaxes.

Table \ref{Table1} displays the TTSs found both in the HBC and in the
Hipparcos Catalogue
(ESA \cite{ESA 1997}). Column 1 gives the star name, while
Columns 2 and 3 indicate its HBC and HIP number, respectively. Entries are
in order of increasing $\alpha$. Column 4 indicates the associated
nebular region, and Column 5 indicates the object type according to the HBC
nomenclature
\footnote{Except for V773 Tau, which is a CTTS with forbidden line emission and
sizable IR excess, and not a WTTS as indicated in HBC.}. A CTTS is
denoted `tt', a WTTS is called `wt', a member of the Herbig Ae/Be (HAeBe)
group is noted `ae', and a SU\ Aurigae-type star is marked `su'.\ A `?'
symbol indicates an uncertain type, while a `*' symbol denotes uncertain
pre-main sequence status. Column 6 gives the median Hipparcos magnitude, and
Column 7 shows the Hipparcos parallax $\pi$ in milliarcsecond (mas), with
standard error $\sigma $ as indicated in Column 8.\ Column 9 indicates the
derived distance in pc for stars with $\pi\geq 2\sigma_\pi$ (marked `:')
and for stars with $\pi\geq 3\sigma_\pi$, with error bars corresponding
to $\pm 1\sigma$. Column 10 gives the flag found in Field H59 (Double and
Multiple Systems Annex flag) of the Hipparcos Catalogue, the meaning of
which we now briefly describe.

For the majority of Hipparcos stars, the astrometric solutions were derived
using a single star model, where the five astrometric parameters are the
equatorial coordinates ($\alpha ,\delta $), the proper motion components
($\mu _{\alpha \ast },\mu _{\delta }$) and the parallax $\pi$. However, detected
non-single stars received different solutions, depending on the
nature of their duplicity. Five different possibilities are noted in the H59
Field by different flags: either the system was resolved into several
components with an assumed linear motion (component solutions, Flag C in
Field H59), or an orbital solution could be computed (Flag O), or the
duplicity was detected by a non-linear motion of the photocentre
(acceleration solution, Flag G), or by the variability of one component,
resulting in a specific motion of the photocentre (VIM solutions: Flag V),
or by an excess scatter of the measurements possibly due to a short-period
variation of the photocentre (stochastic solutions, Flag X). We emphasize
here that the value of the derived parallax depends on the adopted
astrometric model. It may happen, for example, that a single-star model was
given in the Catalogue for a star which is now known to be a spectroscopic
binary. One can then go back to the one-dimensional measurements archived in
the Hipparcos Intermediate Astrometric Data, CD-ROM 5, in order to compute
an orbital solution for this system, resulting in a presumably more accurate
value of the parallax and other astrometric parameters. The reader is
referred to ESA
(\cite{ESA 1997}) for a detailed explanation of astrometric
solutions for non-single stars.

\newcommand{\mcc}[1]{\multicolumn{1}{c}{#1}}
\newcommand{\mcl}[1]{\multicolumn{1}{l}{#1}}

\begin{table*}[ht]
\caption{Parallaxes and binary flags for young
stellar objects in the HBC\label{Table1}}
\begin{tabular}{lrrlcrrrlc}
\hline\hline
\mcl{Star} &
\mcc{HBC} &
\mcc{HIP} &
\mcl{Location} &
\mcc{Ty.} &
\mcc{$\overline{H_p}$} &
\mcc{$\pi$} &
\mcc{$\sigma_\pi$} &
\mcc{$D$}&
\mcc{H59} \\
\mcl{(1)}&
\mcc{(2)}&
\mcc{(3)}&
\mcl{(4)}&
\mcc{(5)}&
\mcc{(6)}&
\mcc{(7)}&
\mcc{(8)}&
\mcc{(9)}&
\mcc{(10)} \\ \hline
\object{V594 Cas}&330&3401&L1291&ae&10.67&3.34&1.63&299:$_{-98}^{+286}$&-\\
\object{XY Per}&349&17890&L1449&ae&9.43&8.33&3.49&120:$_{-35}^{+87}$&C\\
\object{V773 Tau}&367&19762&B209&wt&10.86&9.88&2.71&101$_{-22}^{+39}$&V\\
\object{V410 Tau}&29&20097&B7&wt&10.91&7.31&2.07&137$_{-30}^{+54}$&-\\
\object{BP Tau}&32&20160&L1495&tt&12.12&18.98&4.65&\phantom{0}53$_{-11}^{+17}$&-\\
\object{RY Tau}&34&20387&B214&tt&10.50&7.49&2.18&134$_{-31}^{+54}$&V\\
\object{HDE 283572}&380&20388&B214&su&9.16&7.81&1.30&128$_{-18}^{+26}$&-\\
\object{T Tau}&35&20390&L1546&tt&9.98&5.66&1.58&178$_{-40}^{+67}$&-\\
\object{DF Tau}&36&20777&L1521&tt&12.08&25.72&6.36&\phantom{0}39$_{-8}^{+13}$&V\\
\object{UX Tau A}&43&20990&L1551&wt&11.17&-6.68&4.04&-&V\\
\object{AB Aur}&78&22910&L1517,19&ae&7.08&6.93&0.95&144$_{-17}^{+23}$&-\\
\object{SU Aur}&79&22925&L1517,19&su&9.40&6.58&1.92&152$_{-34}^{+63}$&-\\
\object{UX Ori}&430&23602&L1615?&?&10.70&0.61&2.47&-&V\\
\object{RW Aur}&80&23873&-&tt&10.30&14.18&6.84&\phantom{0}71:$_{-23}^{+65}$&X\\
\object{CO Ori}&84&25540&anon&su&10.78&-1.82&2.80&-&G\\
\object{AB Dor}&435&25647&-&*&7.05&66.92&0.54&\phantom{0}14.9$_{-0.1}^{+0.2}$&G\\
\object{GW Ori}&85&25689&B225&tt&10.00&3.25&1.44&308:$_{-95}^{+245}$&-\\
\object{CQ Tau}&464&26295&-&?&10.63&10.05&2.01&100$_{-17}^{+24}$&-\\
\object{V380 Ori}&164&26327&NGC 1999&ae&10.35&3.72&5.48&-&-\\
\object{BF Ori}&169&26403&L1640,41&?&10.18&-0.67&1.80&-&-\\
\object{HD 250550}&192&28582&L1586&ae&9.55&1.65&1.51&-&-\\
\object{HD 259431}&529&31235&NGC 2247&*&8.70&3.45&1.41&290:$_{-84}^{+200}$&-\\
\object{Z CMa}&243&34042&L1657&ae&9.78&-0.91&2.21&-&V\\
\object{NX Pup}&552&35488&CG1&ae&10.22&1.99&2.38&-&C\\
\object{TW Hya}&568&53911&-&tt&11.07&17.72&2.21&\phantom{0}56$_{-6}^{+8}$&-\\
\object{CR Cha}&244&53691&Cha 1&tt&11.45&6.97&1.87&143$_{-30}^{+52}$&-\\
\object{DI Cha}&245&54365&Cha 1&tt&10.81&4.77&2.82&-&X\\
\object{CU Cha}&246&54413&Cha 1&ae&8.53&5.70&0.76&175$_{-20}^{+27}$&-\\
\object{CV Cha}&247&54738&Cha 1&tt&11.20&3.14&7.39&-&C\\
\object{CW Cha}&589&54744&Cha 1&tt&13.76&3.14&7.39&-&C\\
\object{T Cha}&591&58285&DCld 300.2-16.9&?&11.95&15.06&3.31&\phantom{0}66$_{-12}^{+19}$&-\\
\object{IM Lup}&605&78053&Lup 3&wt&11.72&-4.77&13.79&-&C\\
\object{RU Lup}&251&78094&Lup2&tt&11.17&4.34&3.56&-&-\\
\object{RY Lup}&252&78317&Lup3,4&tt&11.56&9.26&2.83&108$_{-25}^{+48}$&-\\
\object{V856 Sco}&619&79080&Lup 3&ae&7.07&4.81&0.87&208$_{-32}^{+46}$&-\\
\object{V1121 Oph}&270&82323&L162&tt&11.42&10.51&2.77&\phantom{0}95$_{-20}^{+34}$&-\\
\object{AK Sco}&271&82747&anon&tt&9.35&6.89&1.44&145$_{-25}^{+39}$&-\\
\object{IX Oph}&272&83963&B59&?&11.05&-0.26&2.64&-&V\\
\object{FK Ser}&664&89874&L405?&tt&10.75&9.42&6.17&-&C\\
\object{R CrA}&288&93449&NGC6729&ae&11.61&121.75&68.24&-&X\\
\object{V1685 Cyg}&689&100289&anon&ae&10.62&9.25&2.23&108$_{-21}^{+56}$&V\\
\object{BD +41 3731}&693&100628&L897,99&?&9.92&0.41&1.15&-&-\\
\object{HD 200775}&726&103763&NGC7023&ae&7.41&2.33&0.62&429$_{-90}^{+156}$&-\\
\object{BD +46 3471}&310&107983&IC5146&ae&10.22&-0.83&1.47&-&-\\
\object{DI Cep}&315&113269&-&tt&11.50&3.50&2.15&-&-\\
\object{MWC 1080}&317&114995&anon&*&11.56&-6.98&3.33&-&-\\ \hline
\end{tabular}
\end{table*}

Cross-identification of the HBC and Hipparcos stars is in most cases
obvious. Only 4 stars of Table 1 are not cross-identified in the SIMBAD
database. They are the visual pair HIP 54738 and 54744, and the two single
stars HIP 78053 and HIP 114995. The latter two stars are unambiguously
identified as IM Lup and V628 Cas (MWC 1080) by inspection of the
relevant sky regions. The situation is more confused for the HIP 54738 and
54744 pair, which is identified as CCDM J11125 -7644A/B in SIMBAD. Comparing
the Hipparcos Input Catalogue (HIC, Turon et al.
\cite{Turon et al. 1992}) to the Hipparcos Catalogue, it seems
that \object{HIC 54738} was erroneously written down as
\object{HIP 54744}. A careful
examination of the sky atlas and arguments given in Sect. 4 lead us to
propose the identification HIP 54738 = CV Cha and HIP 54744 = CW Cha.
\begin{table*}[ht]
    \caption{Additional young stellar objects connected with the T
associations and used in distance determinations\label{Table2}}
\begin{tabular}{llrlcrrrlc}
\hline\hline
\mcl{Star} &
\mcl{ROSAT} &
\mcc{HIP} &
\mcl{Location} &
\mcc{Ty.} &
\mcc{$\overline{H_p}$} &
\mcc{$\pi$} &
\mcc{$\sigma_\pi$} &
\mcc{$D$}&
\mcc{H59} \\
\mcl{(1)}&
\mcl{(2)}&
\mcc{(3)}&
\mcl{(4)}&
\mcc{(5)}&
\mcc{(6)}&
\mcc{(7)}&
\mcc{(8)}&
\mcc{(9)}&
\mcc{(10)} \\ \hline
\object{BD +11 533}&RX
J0352.4+1223&18117&Tau&wt&10.0&6.55&1.62&$153_{-30}^{+50}$& -\\
\object{HD 284149}&WKS96 6 &19176&Tau&wt&9.71&6.43&1.84&$156_{-35}^{+62}$& -\\
\object{HD 28150}&- &20780&Tau&ae&6.96&8.04&1.53&$124_{-20}^{+29}$& C\\
\object{BD+17 724B}&WKS96 30 &20782&Tau&wt&9.52&7.69&17.39&-& C\\
\object{HD 283798}&WKS96 50 &21852&Tau&wt&9.69&8.68&1.35&$115_{-16}^{+21}$&
-\\
\object{HD 31648}&- &23143&Tau&ae&7.79&7.62&1.18&$131_{-18}^{+24}$& -\\
\object{HD 37061}&- & 26258 &Orion Neb.&?&6.82&2.77&0.88&
$361_{-87}^{+168}$& -\\
\object{HD 97300}&CHXR 42 &54557&Cha 1&ae&9.06&5.33&1.01&$188_{-30}^{+44}$&
-\\
\object{HD 96675}&RX J1105.9-7607&54257&Cha
1&ae&7.74&6.11&0.60&$164_{-15}^{+18}$& -\\
\object{HD 104237}&RX J1200.1-7811&58520&Cha 3
?&ae&6.65&8.61&0.53&$116_{-7}^{+8}$& -\\
\object{HD 102065}&- &57192&DC 300-17&ae&6.64&5.95&0.52&$168_{-14}^{+16}$& -\\
\object{V1027 Sco}&KWS97 Lup 3 40&79081&Lup
3&ae&6.63&4.15&0.83&$241_{-40}^{+60}$& -\\ \hline
\end{tabular}
\end{table*}

In each star forming region (SFR), we searched for additional
pre-main-sequence stars observed by Hipparcos and located in the vicinity of
the stars given in Table 1, in order to improve the precision of the mean
parallaxes. In addition to HBC stars, we thus considered HAeBe stars not
found in the HBC but listed in the Th\'e et al.
(\cite{The et al. 1994}) Catalogue along
with young stars, mainly of WTTS type, which were discovered by the ROSAT
satellite in the vicinity of star-forming regions or which form multiple
systems with stars of Table \ref{Table1}. We restricted ourselves to those
stars with confirmed pre-main sequence status discussed in the series of
papers on ROSAT observations of SFRs
(Neuh\"{a}user \& Brandner \cite{Neuhauser & Brandner 1998},
Krautter et al. \cite{Krautter et al. 1997},
Alcal\'a et al. \cite{Alcala et al. 1995},
Wichmann et al. \cite{Wichmann et al. 1996},
Alcal\'a et al. \cite{Alcala et al. 1997},
Covino et al. \cite{Covino et al. 1997},
Frink et al. \cite{Frink et al. 1998},
Terranegra et al. \cite{Terranegra et al. 1999}).
Table \ref{Table2} (with entries
similar to Table \ref{Table1}) summarizes properties of these additional
pre-main sequence stars.

Note that the Hipparcos data of a large sample of HAeBe stars, containing a
number of likely members of the class in addition to those contained in HBC,
were recently discussed by van den Ancker et al.
(\cite{van den Ancker et al. 1997}); we thus won't
discuss them individually further here but will use them to compute mean
parallaxes of YSO groups.


\section{Parallaxes of TTSs and associated clouds}


CTTSs and WTTSs observed by Hipparcos are among the brightest members of
their respective classes, although they are some of the faintest stars in
the Hipparcos Catalogue. Among the 31 stars with significant parallaxes in
Tables \ref{Table1} and \ref{Table2}, 14 are in the Taurus-Auriga
star-forming region, 1 is located in Orion, 6 are associated with the
Chameleon and 2 with the Lupus star-forming regions, 3 are within the
Scorpius cloud complex, and the other ones are more isolated HAeBe stars and
the nearby CTTS TW Hya.


\subsection{Data quality}


\begin{figure}
\resizebox{\hsize}{!}{\includegraphics{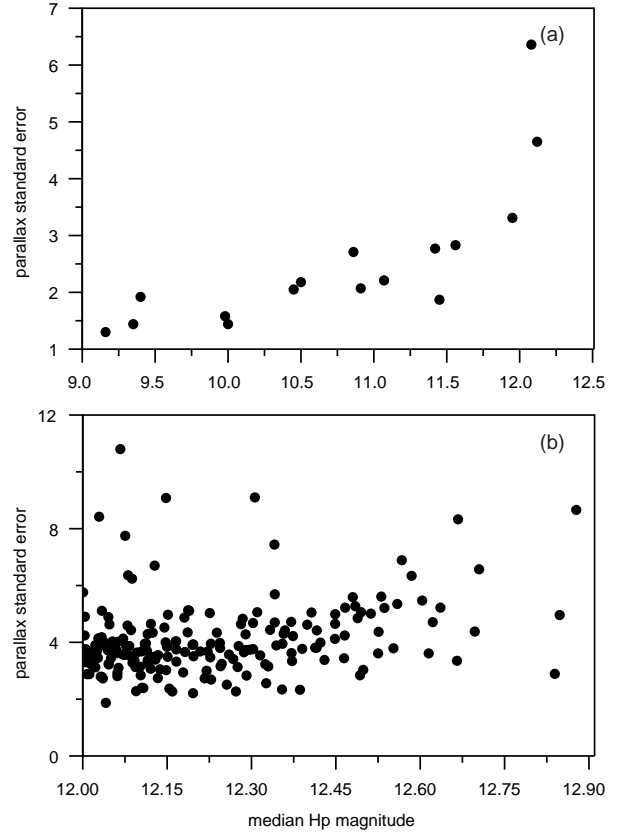}}
  \caption{Panel (a). Relationship between median Hipparcos magnitude
	$\overline{H_{p}}$ and parallax standard error $\protect\sigma $
	for the pre-main sequence stars of \ Table 1. Only positively
	detected stars with $\protect\pi \geq 3\protect\sigma $ are
	plotted here. Panel (b). Same as Panel (a), but for the
	low-luminosity end of the Hipparcos Catalogue. }
  \label{fig1}
\end{figure}

As a first check on data quality, we consider the standard error of the
parallax measured by Hipparcos. While the derived parallax accuracy of the
sample of young stars measured by Hipparcos depends on the median $H_p$
stellar magnitude, it is typically less than about 2~mas, in agreement with
the average accuracy of the Hipparcos Catalogue (cf. Fig. \ref{fig1}). The
faintest stars of this sample, BP Tau and DF Tau, are those two CTTSs whose
distances appear considerably lower than their associated clouds. The
standard errors of their parallaxes is a factor 2 or more larger than
standard errors of other stars in the YSO sample
\footnote{Note that RW Aur has a large parallax error in spite of its relative
brightness. This is presumably due to the strong variability of this CTTS.
Its measured parallax is only slightly above 2$\sigma $ and cannot be
considered significant.}, which should be compared to parallax accuracies
derived for the low-luminosity end of the Hipparcos Catalogue.
Fig. \ref{fig1} also displays (lower panel) parallax standard errors of
the faintest stars with reliable parallaxes found in the Hipparcos Catalogue,
which follows the $\propto 10^{H_p\over 5}$ law expected from photon noise
only. It shows
that DF Tau's parallax standard error is larger by a factor 1.5 than the
standard error of stars with comparable magnitude, while BP Tau does not
stand out in this sample. Except for DF Tau, then, the precision of
parallaxes for the sample of positively detected YSOs appears to be of a
quality consistent with Hipparcos data for stars of comparably low
brightness.


\subsection{Mean distances of T associations}


In each T association, stars were grouped according to their positions and
common motion. Mean parallaxes of these groups were then derived from the
intermediate data (i.e., the abscissae on the Reference Great Circles of the
satellite; cf. ESA \cite{ESA 1997}), using the method developed by
Robichon et al. (\cite{Robichon et al. 1999})
for computing the mean astrometric parameters of
stellar groups. This method explicitly takes into account the fact that the
Hipparcos parameters are correlated within a few square degrees.


\subsubsection{Taurus-Auriga}


\begin{table*}[ht]
\caption{Distance indicators for the Taurus cloud\label{Table3}}
\begin{tabular}{rrrrrrrrc}
\hline\hline
\mcc{HD}&
\mcc{HIP}&
\mcc{$\pi$}&
\mcc{$\sigma _{\pi }$}&
\mcc{$\mu_{\alpha*}$}&
\mcc{$\sigma_{\mu_{\alpha*}}$}&
\mcc{$\mu_{\delta }$}&
\mcc{$\sigma_{\mu_{\delta }}$}&
\mcc{H59} \\ \hline
26154 & 19634 & 6.19 & 1.13 & -15.79 & 1.14 & -25.53 & 0.89 & C \\
28149 & 20789 & 7.86 & 0.75 & -0.35 & 0.75 & -13.25 & 0.70 & - \\
30378 & 22314 & 5.10 & 1.02 & 5.12 & 0.99 & -25.84 & 0.72 & - \\
31293 & 22910 & 6.93 & 0.95 & 1.71 & 1.05 & -24.24 & 0.69 & - \\ \hline
\end{tabular}
\end{table*}

The morphology and distance of the several molecular clouds forming the
Taurus-Auriga complex is summarized by Ungerechts \& Thaddeus (1987).
They adopted a mean distance of 140~pc from previous analyses based on
star counts, photometric distances of reflecting nebulae and reddening
versus distance diagrams of field stars. Several of the young stars observed
by Hipparcos in that region are located in the central part of Taurus, the
distance of which was determined by Racine (1968) and Elias (1978)
from the photometry of a few bright stars associated with nebulosity. By
chance, all the calibrators used by Elias (1978) were observed by
Hipparcos, and in Table \ref{Table3} we give the Hipparcos astrometric
parameters of these objects. The weighted average parallax
${{\sum\left( \pi _{i}/\sigma _{i}^{2}\right)}/
{\sum \left( 1/\sigma _{i}^{2}\right)}}$
and associated standard deviation
$1/\sqrt{\sum \left(1/\sigma_{i}^{2}\right)}$
of these distance indicators is $6.78\pm 0.46$~mas,
corresponding to a distance of 147$_{-9}^{+11}$pc. This rough average
Hipparcos parallax is thus in perfect agreement with previous estimates (cf.
Kenyon et al. \cite{Kenyon et al 1994}).

\def\gof{{\sc gof}}

\begin{table*}[hbt]
\caption{Astrometric parameters of stars in the Taurus-Aurigae complex. For
each group, two mean values, with their standard errors and goodness-of-fit
(\gof), are indicated. The first has been computed using all the stars, while
the second considers only single stars (i.e., stars with an empty H59 field).
\label{Table4}}
\begin{tabular}{lrrrrrrrc}
\hline\hline
\mcl{Name}&
\mcc{HIP}&
\mcc{$\pi$}&
\mcc{$\sigma _{\pi }$}&
\mcc{$\mu _{\alpha*}$}&
\mcc{$\sigma_{\mu_{\alpha*}}$}&
\mcc{$\mu_{\delta}$}&
\mcc{$\sigma_{\mu_{\delta}}$}&
\mcc{H59}\\
\mcl{(1)}&
\mcc{(2)}&
\mcc{(3)}&
\mcc{(4)}&
\mcc{(5)}&
\mcc{(6)}&
\mcc{(7)}&
\mcc{(8)}&
\mcc{(9)}\\ \hline
\multicolumn{9}{c}{Isolated stars} \\ \hline
\object{BD+11 533} & 18117 & 6.55 & 1.62 & 6.37 & 1.79 & -16.60 & 1.50 & - \\
\object{DF Tau} & 20777 & 25.72 & 6.36 & 14.48 & 6.25 & -26.38 & 4.26 & V \\
\object{HD 283798} & 21852 & 8.68 & 1.35 & -0.70 & 1.35 & -20.56 & 1.04 & -
\\ \hline
\multicolumn{9}{c}{Group 1 (L1495 region)} \\
\object{V773 Tau} & 19762 & 9.88 & 2.71 & 0.65 & 2.55 & -24.89 & 1.89 & V \\
\object{V410 Tau} & 20097 & 7.31 & 2.07 & 6.04 & 2.38 & -27.44 & 1.77 & - \\
\object{BP Tau} & 20160 & 18.98 & 4.65 & 5.28 & 5.43 & -33.13 & 3.94 & - \\
\object{RY Tau} & 20387 & 7.49 & 2.18 & 9.08 & 2.62 & -23.05 & 1.89 & V \\
\object{HDE 283572} & 20388 & 7.81 & 1.30 & 7.52 & 1.57 & -27.45 & 1.14 & - \\
\multicolumn{2}{l}{Mean (all stars)}&\multicolumn{2}{c}{$\pi=7.65\pm 0.98$}&\mcl{\gof=1.32}&&&&\\
\multicolumn{2}{l}{Mean (single stars)}&\multicolumn{2}{c}{$\pi=7.97\pm 1.14$}&\mcl{\gof=0.76}&&&&\\
\hline
\multicolumn{9}{c}{Group 2 (Auriga region)} \\
\object{AB Aur} & 22910 & 6.93 & 0.95 & 1.71 & 1.05 & -24.24 & 0.69 &  \\
\object{SU Aur} & 22925 & 6.58 & 1.92 & 0.17 & 2.24 & -21.69 & 1.28 &  \\
\object{HD 31648} & 23143 & 7.62 & 1.18 & 6.25 & 1.22 & -23.80 & 0.91 &  \\
\object{RW Aur} & 23873 & 14.18 & 6.84 & 9.69 & 7.36 & -21.92 & 3.91 & X \\
\multicolumn{2}{l}{Mean (all stars)}&\multicolumn{2}{c}{$\pi=7.08\pm 0.71$} &\mcl{\gof=5.39}&&&&\\
\multicolumn{2}{l}{Mean (single stars)}&\multicolumn{2}{c}{$\pi=7.13\pm 0.75$}&\mcl{\gof=0.63}&&&&\\
\hline
\multicolumn{9}{c}{Group 3 (south Taurus)} \\ \hline
\object{HD 284149} & 19176 & 6.43 & 1.84 & 6.00 & 1.65 & -15.40 & 1.24 & - \\
\object{T Tau} & 20390 & 5.66 & 1.58 & 15.45 & 1.88 & -12.48 & 1.62 & - \\
\object{HD 28150} & 20780 & 8.04 & 1.53 & 2.83 & 2.18 & -17.77 & 1.85 & C\\
\object{BD+17 724B} & 20782 & 7.69 & 17.39 & -5.93 & 25.67 & -33.28 & 20.88
& C\\
\object{UX Tau A} & 20990 & -6.68 & 4.04 & 8.59 & 5.08 & -27.42 & 3.78 & V \\
\multicolumn{2}{l}{Mean (all stars)}&\multicolumn{2}{c}{$\pi=5.66\pm 0.88$}&\mcl{\gof=3.27}&&&&\\
\multicolumn{2}{l}{Mean (single stars)}&\multicolumn{2}{c}{$\pi=5.96\pm 1.20$}&\mcl{\gof=-0.13}&&&&\\
\hline
\multicolumn{9}{c}{All stars} \\ \hline
\multicolumn{2}{l}{Mean (single stars)}&\multicolumn{2}{c}{$\pi=7.21\pm 0.49$}&\mcl{\gof=0.83}&&&&\\
\hline
\end{tabular}
\end{table*}

The 17 stars selected in the Taurus-Auriga region were divided into 3 groups
of respectively 5, 4 and 5 members. The 3 remaining stars HIP~18117, 20777
and 21852 are too isolated to be included in a group. The astrometric
parameters of these stars and their mean values are summarized in Table~\ref
{Table4}. The term \gof{} there refers to the goodness-of-fit of the
astrometric solution (cf. ESA \cite{ESA 1997}, p. 112).

Group 1 contains stars around the cloud L 1495. Their proper motions are
quite similar and reinforce the hypothesis they are part of the same
structure. The mean parallax of these 5 stars is $7.65\pm 0.98$~mas and
$7.97\pm 1.14$~mas when removing the two VIMs HIP~19762 and 20387.
Nevertheless, HIP~20160 (BP Tau), has an individual parallax $18.98\pm 4.65$
mas, larger than these mean values. The mean parallax is
$7.76\pm 1.15$~mas using only the two single stars
HIP~20097 and 20388, which each have an empty H59 field. BP Tau is then at
more than 2$\sigma $ from these values. It is unclear whether the star is a
member of the group with a diverged parallax or is really closer than the
group. In the latter case, it would be an unlikely coincidence that this
star would be just in front of a T association with a similar proper motion
but a distance less than twice as small. We shall discuss BP Tau in some
detail in Sect. 4.

Group 2, in Auriga, contains HIP~22910, 22925, 23143, and 23873. The mean
parallax is $7.08\pm 0.71$~mas and $7.13\pm{0.75}$~mas when removing
HIP~23873 which has a stochastic solution.

Group 3, around T Tau itself, contains HIP 20780, 20782, 20990 to which one
could attach HIP~19176 and 20390 which are a few degrees beside them.
Unfortunately, only HIP~19176 and 20390 have an empty H59 field. HIP 20780 and
20782 form a two-pointing double system (C in H59) in the Hipparcos Catalogue.
HIP~20780 has a reliable solution while HIP~20782 is 3 magnitudes fainter
and has a very uncertain solution. HIP~20990 is a faint VIM with an
unreliable solution. The mean parallax of these 5 stars is $5.66\pm 0.88$
mas. It is $5.80\pm 0.90$~mas using the three reliable stars HIP~19176,
20390 and 20780 and $5.96\pm 1.20$~mas using only the two single stars
HIP~19176 and 20390.

Putting the 8 single stars (H59 empty) together, we obtain a mean parallax
of the Taurus-Auriga complex of $7.21\pm 0.49$~mas corresponding to a
distance of $139_{-9}^{+10}$~pc. Using the most reliable parallax value
for each group (those obtained with the single stars only), the three groups
are respectively at $125_{-16}^{+21}$, $140_{-13}^{+16}$ and
$168_{-28}^{+42} $~pc. These values, although statistically in agreement,
could reflect real distance differences of about the angular size of
the complex.


\subsubsection{Orion}\label{orion}


Orion is a very large complex of molecular clouds with several distinct
regions (cf. Maddalena et al. \cite{Maddalena et al. 1986}).
The complex has practically no
tangential reflex solar motion so that YSOs are impossible to separate from
field stars using astrometric parameters. The only star with detected
(marginally significant) parallax in the Table \ref{Table1} sample is GW
Ori. It is associated with molecular clouds surrounding the HII region
excited by the O8 star $\lambda$ Ori, HIP~26207, at a distance of
$324_{-65}^{+109}$~pc; and the derived distance of GW Ori is consistent with
this value.\ The star CO Ori is also in the same vicinity, but it is fainter
with parallax standard error comparable to its expected parallax.

In addition to the stars listed in Table \ref{Table1}, 15 presumably young
Orion stars are found in the Hipparcos Catalogue: the HAeBe stars HIP~24552,
25299, 25546, 26594, 26752 and 27059 (from van den Ancker et al. 1997), and
9 stars detected by ROSAT in the Orion Nebula cluster
(Gagne et al. \cite{Gagne et al 1995}) --
HIP~26220, 26221, 26224, 26233, 26234, 26235, 26237, 26257 and
26258. Only the last of these stars has a (marginally) significant
parallax. Two other Orion stars detected by Rosat, HIP~26081 and 26926 are
foreground stars according to their Hipparcos parallaxes: respectively
$59.45\pm 3.88$~mas and $19.94\pm 0.83$~mas.
The Orion nebula cluster is probably not a bound cluster but part of 
a 80~pc long structure connected with the Orion OB1 association
(Tian et al. \cite{Tian et al. 1996}). Among the
Hipparcos-detected stars in this cluster, only HIP~26258 has an empty H59
field. HIP~26220, 26221 and 26224 are three components of a quadruple
system, HIP~26233, 26234, 26235 and 26257 are also flagged C in H59, while
HIP~26237 has a stochastic solution. No mean parallax could therefore be
determined for these objects. HAeBe stars outside the Orion nebula cluster
are more isolated and cannot be grouped to compute mean parallaxes.

\begin{table}[ht]
\caption{Distance indicators for the Orion R1 and R2 associations (Racine
1968)\label{TabOri}}
\begin{tabular}{llll}
\hline\hline
HD/BD & HIP & $\pi $ & $\sigma_\pi$ \\ \hline
36540 & 25954 & 1.79 & 1.19 \\
-06 1253 & 26327 & 3.72 & 5.48 \\
37674 & 26683 & 3.08 & 1.07 \\
37776 & 26742 & 1.96 & 0.98 \\
38087 & 26939 & 5.02 & 1.89 \\ \hline
\end{tabular}
\end{table}

We can nevertheless derive a new, post-Hipparcos estimate of the distance to
the Orion A and B clouds by considering the distance indicators of
Racine (\cite{Racine 1968}),
which are members of the Orion R1 and R2 associations. Among
the 8 stars studied by Racine, 5 were observed by Hipparcos. Their
parallaxes are given in Table \ref{TabOri}. All these stars are single with
empty H59 and H61 fields, except for HD 38087, a double with published
component solution. The mean weighted parallax of the Orion distance
indicators, to which we add the star HIP 26258 discussed above, is $2.63\pm
0.49$~mas. This corresponds to a distance of $381_{-59}^{+86}$~pc, to be
compared to the value of $600\pm 50$~pc derived by Racine.


\subsubsection{Chamaeleon}


The Chamaeleon, Lupus, and Scorpius clouds have spatial velocities close to
that of the Sco OB2 (Scorpius-Centaurus-Lupus-Crux) association,
around $(U,V,W)$ $=(0,-12,0)\pm (5,5,5)$~km~s$^{-1}$. 
The histories of these SFRs are
certainly connected even if their present ages, motions and distances are
not exactly the same.

The distance of Chamaeleon clouds has been a topic of controversy for a long
time; earlier estimates ranged from 115 to 215~pc for Cha 1, and from 115 to
400~pc for Cha 2. From a recent re-investigation of the reddening with
distance over a large area around the clouds,
Whittet et al. (\cite{Whittet et al 1997})
conclude that the most probable distance of Cha 1 is $160\pm 15$~pc, and
that of Cha 2 is $178\pm 18$~pc. A similar analysis using Hipparcos
distances gives essentially the same result
(Knude \& H\o g \cite{Knude & Hog 1998}). 

We can use 10 stars to compute mean parallaxes of groups in the Chamaeleon
region. In Cha 1 there are four CTTSs: HIP~53691, 54365 and the pair HIP
54744 and~54738. The two Herbig Ae/Be stars HIP~54413 and HIP~54557 are
usually associated with Cha 1 because of their reflecting nebulosities.
Inclusion of HIP~54257 is more speculative, since it is a B star not known
as a Herbig star but detected by ROSAT. On the basis of its Hipparcos proper
motion and parallax,
Terranegra et al. (\cite{Terranegra et al. 1999}) believe that it is a
probable member although Gry et al. (1998) place it just behind the Cha
1 complex. The mean parallax of these 7 stars is $5.86\pm 0.45$~mas, whereas
it is $5.96\pm 0.45$~mas when considering only the four stars with an empty
H59 field, corresponding to our best estimate of 168$_{-12}^{+14}$~pc for
the Cha 1 distance. Note that the group mean parallax is only $5.71\pm 0.62$
mas when also removing the possibly dubious member HIP~54257, which rules
out that this star is located behind the cloud, since its individual
parallax is $6.11\pm 0.60$~mas.

The other subgroups of the Chamaeleon region have few Hipparcos stars. There
is the WTTS HIP~58285 in Cha 3, the B star HIP~57192 just in the head of DC
300--17, and the HAeBe star HIP~58520 at less than 1 degree from the head of
DC 300-17 but connected to Cha 3 in the literature. In addition, the B star
HIP~57192 is placed just behind the DC 300-17 cloud by
Gry et al. (\cite{Gry et al. 1998}).
Its parallax is $5.95\pm 0.52$~mas, in perfect agreement with the mean
parallax of Cha 1. HIP~58520 ($\pi =8.61\pm 0.53$~mas) is definitely
closer than the Cha 1 cloud. HIP~58285 is faint with a large parallax error
($\pi =15.06\pm 3.31$) but its astrometric parameters are closer to those of
HIP~58520 than to those of HIP~57192. HIP~58520 and 58285 are members of a
presumed moving group of young stars described in
Frink et al. (\cite{Frink et al. 1998})
and
Terranegra et al. (\cite{Terranegra et al. 1999}).
Two other Hipparcos WTTSs belong to this
proposed group, the reality of which is difficult to demonstrate. Its
spatial velocity is small and is noticed mainly because of its reflex motion
with regard to the Sun. Note that one can find many stars with the same
motion in the Hipparcos Catalogue, in a distance range of 50-200~pc, which
were not detected by Rosat. While some of them may form a moving group with
the WTTSs mentioned above, it is also conceivable that this apparent group
is an artifact caused by the limiting magnitudes of Hipparcos and ROSAT.

To summarize this section, the only firm evidence concerning the Chamaeleon
complex is that Cha 1 is at a distance of about 170~pc, in agreement with
recent estimates by
Whittet et al. (\cite{Whittet et al 1997}) and
Knude \& H\o g (\cite{Knude & Hog 1998}).


\subsubsection{Lupus}


Re-assessing association membership from Hipparcos results,
de Zeeuw et al. (\cite{de Zeeuw et al. 1999})
find a mean distance of 140~pc for the Upper Centaurus Lupus
association.\ From the angular extent of the association (about
$27^{\circ}$), and assuming that it is nearly spherical, the distances of
individual members are expected to range from about 110 to 190~pc. Two
positively detected stars of our sample (RY Lup and V856 Sco) are apparently
associated with clouds of the Lupus star-forming region, and their distances
are compatible with the above range.

Six stars connected with the Lupus complex can be used for computing mean
parallaxes: HIP~77157, 78094, 78317, 79080, 79081 and 78053. HIP~77157,
78094 and 78317 are TTSs in Lupus 1, 2 and 4 respectively. HIP 79080 and
79081 are the brightest stars of a quadruple system (at least) in Lup 3.
HIP~79080 is a Herbig Ae star and 79081 a peculiar B star. HIP~78053 is a
WTTS in Lupus 3 with a poor Hipparcos solution and a flag C in field H59. We
don't include six other WTTSs discovered by ROSAT
(Krautter et al. \cite{Krautter et al.
1997} and
Wichmann et al. \cite{Wichmann et al. 1997}) and discussed by
Neuh\"{a}user \& Brandner (\cite{Neuhauser & Brandner 1998})
because they have proper motions not exactly in agreement
with CTTSs, so that it is quite impossible to decide whether they are
members of the T association, the OB association or are young field stars
instead.

The mean parallax computed for the 5 stars with an empty H59 field is
$4.77\pm 0.61$~mas, a small value compared to the previous estimates quoted
above. However, it results mainly from the brightest stars HIP~79080 and
79081, which have small errors on their parallaxes. If we remove these stars
and consider only HIP~77157, 78094 and 78317, we obtain a larger value of
$6.79\pm 1.50$~mas -- in agreement with previous determinations of the Lupus
SFR distance, but with a large uncertainty -- and a value of $4.38\pm 0.67$
mas for HIP~79080 and 79081. The explanation could be either that HIP~79080
and 79081 are not members of Lupus 3 or that this cloud is farther away than
the other associations.


\subsubsection{Scorpius}


This is another association studied by de Zeeuw et al. (1999), who find
a mean distance of 145~pc, in agreement with previous work by
Racine (\cite{Racine 1968}).
The parallax distribution width is approximately 1~mas,
corresponding to distances from 127 to 170~pc. While the parallax of AK Sco
is consistent with this result, the fainter CTTS V1121 Oph may be located
somewhat closer to us. Note that the quadruple system $\varrho $ Oph
\footnote{$\varrho $ Oph AB = HIP~80473, $\varrho $ Oph C = HIP~80474,
$\varrho $ Oph D = HIP~80461}, associated with the dense parts of the cloud
where vigorous star formation is currently taking place, is at a distance of
$128_{-10}^{+12}$~pc, i.e., on the front side of the Upper Scorpius
association.


\subsubsection{The \object{TW Hya} association}


TW Hya, the closest CTTS, has long been considered to be isolated but is now
known to be part of the closest T association (cf.
Reza et al. \cite{Reza et al. 1989},
Gregorio-Hetem et al. \cite{Gregorio-Hetem et al. 1992},
Kastner et al. \cite{Kastner et al. 1997},
Jensen et al. \cite{Jensen et al 1998},
Sterzik et al. \cite{Sterzik et al 1999} and
Webb et al. \cite{Webb et al. 1999}). The
most recent list of members reports 13 pre-main sequence systems in the
vicinity of TW Hya for a total of at least 20 objects ranging from an A star
to a possible brown dwarf (Sterzik et al. \cite{Sterzik et al 1999}).

\begin{table}[ht]
\caption{Hipparcos-detected members of the TW Hya association)\label{Table6}}
\begin{tabular}{lrrr}
\hline\hline
\mcl{Name}&
\mcc{HIP}&
\mcc{$\pi$}&
\mcc{$\sigma_\pi$}\\ \hline
\object{HD 98800} & 55505 & 21.43 & 2.86 \\
\object{CD -36 7429} & 57589 & 19.87 & 2.38 \\
\object{HD 109573} & 61498 & 14.91 & 0.75 \\ \hline
\end{tabular}
\end{table}

Four of these objects are in the Hipparcos Catalogue: HIP~53911 (TW Hya
itself) and the three objects given in Table \ref{Table6}. Their parallaxes
lead to a depth of 20~pc, comparable to the angular size of the association.
We didn't compute a mean parallax because the method is not suited to such a
nearby group where the depth is not small in comparison to the formal
parallax errors.

\subsubsection{Conclusion}

In most cases, we confirmed that the Hipparcos distances to young solar-type
stars are comparable to the distances of molecular clouds and/or OB stars
with which they appear associated, as anticipated from a large body of
earlier work. There are, however, two discrepant individual cases: BP Tau
and DF Tau. These are among the most investigated CTTSs, so one may argue
that a good part of what we know about the T Tauri phenomenon is based on
these stars. On the other hand, there are a few CTTSs that are found
outside of molecular clouds, such as RW Aur or, as seen above, TW Hya.
Perhaps such cases are not so rare? Before we start to speculate, it appears
quite important to convince ourselves that the Hipparcos results for BP Tau
and DF Tau are valid. A possible bias can be due to undetected binarity,
and we shall examine in the next section the evidence for binarity in
Hipparcos data for YSOs.


\section{Binarity of YSOs in the Hipparcos Catalogue}


A few known young binary systems (FK Ser, etc.) were included in the
Hipparcos observation program, but most of the discoveries of close T Tauri
systems result from recent progress in high-resolution interferometric and
imaging techniques, after the Hipparcos launch. This fast moving research
field opens the exciting prospect of understanding the molecular core
fragmentation process and the relationship between the formation of binaries
and disk (cf. Mathieu \cite{Mathieu 1994}).

The theoretical angular resolution of Hipparcos is $0.45$ arcsec, but double
stars with a moderate brightness ratio ($\Delta H_p<\approx 2$) separated
by 0.10-0.13 arcsec already produce a measurable broadening of the diffraction peak
(Lindegren 1997). Unresolved systems with separation below this limit
(and below about 0.3 arcsec for larger brightness ratios) were seen by
Hipparcos as single point sources located at the photocentre of the system.
Both orbital motion and light variability of at least one of the system's
components can lead to deviations of the photocentre's path from its
expected uniform motion, giving rise to the suspicion that the star is an
unresolved binary. As explained above, if an astrometric solution for a
binary system could be derived by the Hipparcos data reduction teams, it is
indicated in the Hipparcos Catalogue (Field H59, cf. Column 10 of Table \ref
{Table1}). Table \ref{Table1} shows that Hipparcos observations resulted in
several detections of astrometric binaries in the small YSO sample that was
observed.

We went back to the intermediate astrometric data of these objects and of
other young stars for which the astrometric solutions derived by the
Hipparcos teams did not take into account newly discovered information,
notably as far as binarity is concerned. We tried on a star-by-star basis to
explore alternative astrometric solutions, and we discuss below those
systems where an improvement has been obtained, in order of increasing
$\alpha$ for each type of double star distinguished in Hipparcos
astrometric solutions.


\subsection{Variability-induced movers}


Since TTSs are variable stars, it is not surprising to find that several of
them are called Variability-Induced Movers (VIMs, Field H59=V), for which
Wielen (\cite{Wielen 1996}) derived astrometric solutions based on the
assumption
that one star in the binary system is variable, while the other one is
constant.\ Since the photocentre varied back and forth between the two
components, both the position angle $\Theta _{C}$ (with its standard
deviation $\sigma_{\Theta_{C}}$) and a lower limit $\rho_{\min }$ of the
separation can be found in this way. The results for the 8 VIMs among our
TTS sample are given in Table \ref{Table5}, and we discuss each of these
objects briefly in turn.

\begin{table}[h]
    \caption{VIM solutions for Young Stellar Objects\label{Table5}}
\begin{tabular}{lrr}
\hline\hline
\mcl{Star} &
\mcc{$\Theta _{C}\pm \sigma _{\Theta _{C}}(^{\circ })$}&
\mcc{$\rho _{\min }$ (mas)}\\
\mcl{(1)} & \mcc{(2)} & \mcc{(3)} \\ \hline
{V773 Tau} & {118.08 $\pm $ \phantom{0}9.54} & {119.4} \\
{RY\ Tau} & {316.61 $\pm $ 37.59} & {19.5} \\
{DF\ Tau} & {307.08 $\pm $ 17.70} & {80.2} \\
{UX\ Tau A} & {253.90 $\pm $ \phantom{0}5.53} & {204.7} \\
{UX\ Ori} & {257.42 $\pm $ 18.42} & {21.8} \\
{Z\ CMa} & {135.30 $\pm $ 20.27} & {80.1} \\
{IX Oph} & {244.59 $\pm $ 13.65} & {56.9} \\
{V1685 Cyg} & {22.94 $\pm $ 11.71} & {140.6} \\ \hline
\end{tabular}
\end{table}


\subsubsection{\object{V773 Tau}}


This is a triple system made up of a double line spectroscopic binary star
(Welty \cite{Welty 1995}) and a third component which has been resolved at
several
optical and IR wavelengths (cf. Ghez et al. \cite{Ghez et al. 1997a} and
references
therein). The position angle of V773 Tau C ranges from 295$^{\circ }$ in
1990 to 320$^{\circ }$ in 1994, and the separation is about 115~mas until
1993, then decreasing to about 60~mas from 1993 to 1995. This is in good
agreement with the VIM solution given in Table \ref{Table4}. Note that the
position angle $\Theta _{C}$ refers to the constant component of the binary
(as assumed in the VIM formalism) with respect to the variable component,
while it refers to the primary component in Ghez et al. (\cite{Ghez et al.
1997a}). The
phase shift of $\pi $ between the two position angle values thus confirms
that V773 Tau C is more variable than V773 Tau AB, as observed by
Ghez et al. (\cite{Ghez et al. 1997a}). It is therefore likely that the two
flares recorded by
Hipparcos in its photometric database originate from V773 Tau C.

There is an unmodelled scatter in the astrometric data residuals of the
Hipparcos
solution, which could be due to the motion of the photocentre of the
two-lined spectroscopic binary. However, if one uses the luminosities of the
components from Ghez et al. (\cite{Ghez et al. 1997a}) and the~masses
($\times \sin i$)
given by Welty (\cite{Welty 1995}), the semi-major axis for the motion of the
photocentre is $\approx 5$\% of the semi-major axis of the relative orbit,
i.e., around 0.15~mas. This is too small for detection by
Hipparcos, and even if the secondary had a much smaller luminosity, the
effect would be negligible compared to the VIM effect ($\approx 58$~mas
variation of the photocentre between the minimum and maximum luminosity of
the system).

Recently, Lestrade et al. (\cite{Lestrade et al. 1999}) determined the
astrometric parameters
of V773 Tau using high precision VLBI astrometry, and found a parallax of
$6.74\pm 0.25$~mas. The derived parallax and proper motion agree
within $2\sigma$ with Hipparcos results. The fact that the
position angle of V773 Tau C changed during the mission has an influence on
the astrometric parameters: constraining the position angle to vary linearly
with time during the 3-year mission with $\theta =6.33t+118^{\circ}$ gives
a parallax of $8.74\pm 3.19$~mas, closer to the VLBI value but with a
large uncertainty.


\subsubsection{\object{RY Tau}}


This star has long been suspected of being a binary (Herbig \& Bell
1988) based on apparent changes in the stellar radial velocity, and
Hipparcos confirmed this suspicion.

A new VIM solution was computed, using all available intermediate data, and
slightly improving the published solution. The position angle is 
$304\pm 34^{\circ }$ and the lower limit on the separation is 23.6~mas. This is
compatible with the lack of detection of this system in current high angular
resolution observations. The projected minimum distance between the two
components is 3.27 AU.


\subsubsection{\object{DF Tau}}


The binarity of DF Tau was first detected in a lunar occultation (\cite
{Chen et al. 1990}), and was observed on several occasions since then.
Available data are summarized by Ghez et al. (\cite{Ghez et al. 1997a}),
who find that the
position angle decreases from $328^{\circ }$ in 1990 to $297^{\circ }$ in
1995, while the separation stays approximately constant over this time, at
about 90~mas. The position angle from the Hipparcos VIM solution is
consistent with these values and indicates that the primary is the variable
component. This is also suggested by
Ghez et al. (\cite{Ghez et al. 1997a}), who note that
both components display infrared excess but that the UV excess, which
signals accretion activity, is much stronger in the primary than in the
secondary. The VIM formalism (used to derive DF Tau's parallax in the
Hipparcos Catalogue) assumes uniform orbital motion, which is obviously not
the case here. We tried to improve the astrometric solution in two ways.

\begin{enumerate}
\item  Assuming that the VIM formalism is correct, i.e. that the
variability-induced photocentre motion dominates the orbital motion, we
derived a new VIM solution using a linear change of the position angle during the
mission duration. The parallax in this case is $18.52\pm 8.61$~mas, which
confirms that the derived value depends critically on the assumed
astrometric model.

\item  We then tried to obtain a new astrometric solution by assuming that
the observed motion is orbital, i.e., by neglecting the variability-induced
photocentre motion. In support of this procedure, one can argue that the $H_p$
magnitude should always be dominated by the primary, since the contrast
between the two components ranges from 3.4 in the HST F439W B-like filter to
1.9 in the HST F555W V-like filter, and the primary is also the most active
component, as discussed above. We thus used the (admittedly preliminary)
orbital parameters determined by
Thi\'ebaut et al. (\cite{Thiebaut et al. 1995}) to compute a
new astrometric solution, and found a parallax value of $14.06\pm 9.06$
mas.
\end{enumerate}

Obviously, a solution combining both orbital motion and VIM is necessary for
this object, but this is impossible without more precise information,
notably about the orbital parameters. None of the above derived parallaxes
is significant, and the published parallax should obviously be considered
with caution. Can we conclude that a location of DF Tau outside of the
Taurus cloud is ruled out by our analysis? Not definitely. We have merely
shown that the derived parallax in the Hipparcos Catalogue is probably not
significant, and that its standard error is so large that DF Tau's weight
in mean parallax derivations is very low. The only firm conclusion that we
can draw is that, as a group including DF Tau, the TTSs associated with
Taurus are indeed located at the cloud's distance; as for DF Tau, its
location remains uncertain.


\subsubsection{\object{UX Tau A}}


Prior to Hipparcos launch, this object was a known triple system (\cite
{Jones & Herbig 1979}). UX Tau A and B components are WTTSs, while UX Tau C
is a
low-mass object with H$\alpha $ emission. The astrometric companion found by
Hipparcos is most likely the B component. Table \ref{UXTauTab} summarizes
the current position angles and separations of components B and C with
respect to A.

\begin{table}[ht]
\caption{The triple system UX Tau \label{UXTauTab}}
\begin{tabular}{crr}
\hline\hline
UX Tau & \mcc{p.a. ($^{\circ }$)}&\mcc{$\rho $ ('')}\\ \hline
B & 269 & 5.9 \\
C & 181 & 2.7 \\ \hline
\end{tabular}
\end{table}


\subsubsection{\object{UX Ori}}


Binarity of this star was first detected by Hipparcos. The small minimum
separation found in this VIM solution may explain why this star has not been
detected in IR with the shift-and-add technique (Pirzkal et al. \cite{Pirzkal
et al. 1997}). It has been argued that the observed variability of this star
and of BF Ori (also suspected of binarity by Hipparcos) is due to violent
comet-like activity
(Grinin et al. \cite{Grinin et al. 1994},
de Winter et al. \cite{de Winter et al. 1999}).


\subsubsection{\object{Z CMa}}


This star was already known as a binary with position angle around
$120^{\circ}$ and separation 100~mas (Koresko et al.
\cite{Koresko et al. 1991},
Leinert et al. \cite{Leinert et al. 1997}). The Hipparcos solution is in
agreement with these solutions.
In the optical range, the variable component is the primary.


\subsubsection{\object{IX Oph}}


The evolutionary status of this F star is not entirely clear, and it appears
to have attracted little observational attention so far. The detection by
Hipparcos of its binary nature is a new development. An improved VIM
solution gives a position angle $243\pm 10^{\circ }$ and minimum
separation of 46.84~mas.


\subsubsection{\object{V1685 Cyg}}


Although the minimum separation, according to Hipparcos, is rather large,
the binarity of this Herbig B2,3e star was detected neither in high angular
resolution IR observations (Pirzkal et al. \cite{Pirzkal et al. 1997}), nor by
speckle-interferome\-try
(Leinert et al. \cite{Leinert et al. 1997}). However, the region
around this star is a small stellar cluster with a molecular outflow
oriented north-south
(Palla et al. \cite{Palla et al. 1995}), with V1318 Cyg and V1686
Cyg located south of BD +40 4124, which could explain why a VIM solution was
found with a position angle $20\pm 11^{\circ}$.


\subsection{Component solutions}



\subsubsection{\object{XY Per}}


This binary has a $1.331\pm 0.01$ arcsec separation, 76.3$^{\circ }$
position angle with 0.88 $H_{p}$ magnitude difference. This is consistent
with the results of
Pirzkal et al. (\cite{Pirzkal et al. 1997}), who find respectively 1.2
arcsec and 255$^{\circ }$. Hipparcos thus resolves the 180$^{\circ }$
ambiguity noted in
Pirzkal et al. (\cite{Pirzkal et al. 1997}) and caused by the nearly equal
brightness of the components.


\subsubsection{\object{NX Pup}}


This is a HAeBe star whose binary nature was discovered with the HST Fine
Guidance Sensor giving $\rho=0.126\pm .007$ arcsec, 
$\theta =63.4\pm 1.0$, with a $0.64\pm .07$ magnitude difference
(Bernacca et al. \cite{Bernacca et al 1993}).
Hipparcos found consistent but less precise estimations,
respectively $0.140\pm .026$ arcsec, $\theta =74^{\circ }$, and $\Delta
H_{p}=0.44\pm 1.10$ mag.


\subsubsection{\object{CV Cha}}


CV Cha and CW Cha are the two components of an optical binary T Tauri
system with separation equal to 11.4~arcsec and p.a. $105^{\circ}$
(Reipurth \& Zinnecker \cite{Reipurth & Zinnecker 1993}).
HIP 54744 is identified as CCDM J11125-7644A in SIMBAD, 
while HIP 54738 is identified as CCDM J11125-7644B. The
component solution derived in the Hipparcos Catalogue gives a separation equal
to 8.48 arcsec and a position angle of $275^{\circ}$. The solution quality
is given as poor (`C'), and the Hipparcos solution is not consistent with
the images obtained by Reipurth \& Zinnecker
(\cite{Reipurth & Zinnecker 1993}). However, we note
that the Hipparcos-derived position angle would be consistent with
observations, if HIP 54738 were in fact CV Cha and if HIP 54744 were CW
Cha. Given the weakness of CW Cha, the separation derived by Hipparcos is
likely to be inaccurate. As discussed also in Sect. 2, we conclude that
there is a misidentification in the Hipparcos Catalogue, and we believe that
HIP 54738 = CCDM J11125 -7644A = CV Cha, while HIP 54744 = CCDM J11125
-7644B = CW Cha. A single star astrometric solution for CV\ Cha gives the
following results: $\pi=7.60\pm 2.10$~mas, $\mu_{\alpha }\cos\delta
=-21.00\pm 2.19$~mas/yr, and $\mu _{\delta }=0.38\pm 1.94$~mas/yr
(Falin, priv. comm.). Results
given for HIP 54744 in the Hipparcos Catalogue should be discarded.


\subsubsection{\object{FK Ser}}


This visual binary was found by Herbig to be a possible post-T Tauri system
(Herbig \cite{Herbig 1973}). Hipparcos measured a separation of $1.118\pm 0.025$
arcsec, and the position angle of the secondary is 14$^{\circ }.$ These
values can be compared to those given by Herbig for the date 1972.5:
separation of 1.32 arcsec, p.a. 11.5$^{\circ}$.


\subsection{Acceleration solutions}


An acceleration solution, using either a quadratic or cubic motion with
respect to time, was applied to all stars not having a `component solution',
and only the stars with significant non-linear terms were retained. The
acceleration effect may be interpreted as the signature of binaries with an
intermediate period (more than about 10 years).


\subsubsection{\object{CO Ori}}


This star has been detected as a binary by Reipurth \& Zinnecker (1993),
who mention a 280$^{\circ}$ position angle and 2.0 arcsec separation.
Given the distance of Orion and the very long period, it is unlikely that
the acceleration term may be significant, so that the detected variation of
the photocentre with time is probably an artifact due to the configuration
of the system and the scanning law of the satellite.

If a stochastic solution (see below) is applied instead of an acceleration
solution, the cosmic error is $4.31\pm 1.34$~mas, i.e., with the same
significance as a Gaussian $2\sigma $ level. From this cosmic error a
magnitude difference $\Delta H_{p}=3.19\pm 0.3$ mag is estimated, consistent
with the 0.07 flux ratio in the Gunn $z$ band. If a VIM solution is
computed, the astrometric elements of the VIM motion are not significant at
more than a $1\sigma $ level, but it should be noticed that the position
angle found, $301\pm 30^{\circ}$, is also consistent with the
Reipurth \& Zinnecker \cite{Reipurth & Zinnecker 1993} observation.


\subsubsection{\object{AB Dor}}


The combination of Hipparcos and VLBI data allowed the determination of a
dynamical~mass of about $0.09$ solar~mass for the low-mass companion of this
ZAMS star (Guirado et al. \cite{Guirado et al. 1997}). The Hipparcos data
cover only a
small fraction of the period, but the curvature of the photocentre motion
was nevertheless clearly detected.


\subsection{Stochastic solutions}


These solutions were applied as a last resort during the Hipparcos data
reduction, when all other solutions failed to give an adequate astrometric
fit. A so-called cosmic error $\epsilon$ was added to the abscissae
standard error, representing the unmodelled photocentre variations. Although
the photocentre displacement may be due to short-period astrometric binaries
(e.g. HIP~39903, Arenou \cite{Arenou 1998}), it may also be caused by
undetected
long-period binaries, with separation of a few arcseconds. It may also be
that a stochastic solution simply reflects bad abscissae measurements,
without any binarity indication.

For resolved binary systems, there is a correlation -- depending weakly on
separation -- between the magnitude difference of the two components $\Delta
H_{p}$ and the cosmic error $\epsilon$ that would result if a stochastic
solution was computed instead of a component solution (Arenou \cite{Arenou
1997}).
Using all Hipparcos component solutions with separation, e.g., between 1.3
and 1.5 arcsec, the magnitude difference $\Delta H_{p}$ can be calibrated
against the cosmic error $\epsilon$, leading to
\[\Delta H_p\approx (-0.90\pm 0.01)\ln\epsilon +4.50\pm 0.04\]
This relationship allows us to estimate the magnitude difference between
components when Hipparcos does not resolve a binary system.


\subsubsection{\object{RW Aur}}


This is a triple star (Table \ref{RWAurTab};
cf. Ghez et al. \cite{Ghez et al. 1997a})
with the A component separated by 1.4 arcsec from the BC binary (0.12 arcsec
separation), which probably perturbed the Hipparcos observations.

\begin{table}[h]
    \caption{The triple system RW Aur \label{RWAurTab}}
\begin{tabular}{lll}
\hline\hline
RW\ Aur & p.a. ($^{\circ }$) & $\rho $ ('') \\ \hline
B (wrt A) & 255 & 1.417 \\
C (wrt B) & 111 & 0.120 \\ \hline
\end{tabular}
\end{table}

No significant VIM solution can be found, but it is sufficient to reject the
bad abscissae to obtain a better astrometric solution $\pi =7.98\pm 3.15$
mas, $\mu _{\alpha \ast }=3.26\pm 3.44$ and $\mu _{\delta }=-23.03\pm 2.00$
mas/y. This justifies the use of this star for computing a mean distance of
the Taurus-Auriga region. Estimated magnitude difference between BC and A is
$\Delta H_{p}=2.08\pm 0.07$ mag, not far from the $2.3\pm 0.08$ bolometric
magnitude difference given by Ghez et al.
(\cite{Ghez et al. 1997a}).


\subsubsection{\object{DI Cha}}


Although the binarity of this object is not detected, e.g., in the infrared
DENIS survey
(Cambr\'esy et al. \cite{Cambresy et al. 1998}), it is clear from
Reipurth \& Zinnecker (\cite{Reipurth & Zinnecker 1993}),
Chelli et al. (\cite{Chelli et al. 1995}), and
Ghez et al. (\cite{Ghez et al. 1997b})
that this is a binary of separation $4.9\pm 0.3$ arcsec and position angle
$202\pm 3^{\circ }$.

However, the binarity has probably not perturbed the Hipparcos astrometry
too much. Indeed, if this star was reduced as a single star, its parallax
would be $5.16\pm 1.54$~mas, close to the stochastic solution, although with
a 1.21 normalized $\chi^2$.

Using the same method as above, the calibrated relation between cosmic error
and magnitude difference is also valid for separation between 4 and 6
arcsec, so that the magnitude difference for DI Cha components is $\Delta
H_{p}=2.54\pm 0.09$ mag. Note that the primary is redder than the secondary
with a difference of about 4 mag in the K band
(Chelli et al. \cite{Chelli et al. 1995}).


\subsubsection{\object{R CrA}}


The Corona Australis molecular complex is at a distance of about 130~pc
(Marraco \& Rydgren \cite{Marraco & Rydgren 1981}),
which is corroborated by the parallax
$7.35\pm 1.15$~mas of HD 176386 (HIP 93425).
R CrA is surrounded by several other
YSOs
(Wilking et al. \cite{Wilking et al. 1997}), which probably explains why
the Hipparcos
observations have been perturbed, leading to a stochastic solution. Although
the error bar on the parallax prevents any safe distance to be derived, it
is clear that the Hipparcos astrometric solution for this star should be
completely discarded.

Once a VIM solution was attempted, the parallax shif\-ted from $121\pm 68.24$
to $36.7\pm 91$~mas. Even constraining the parallax to the expected parallax
of Corona Australis does not give a satisfactory solution (in the sense of
obtaining significant values of astrometric or orbital parameters). We
conclude that the astrometric intermediate data are clearly useless for this
star.


\subsection{Other astrometric solutions}



\subsubsection{\object{V410 Tau}}


This is a triple system (Table \ref{V410Tab},
Ghez et al. \cite{Ghez et al. 1997a}),
undetected by Hipparcos, apart from the `suspected non-single' flag in the
Hipparcos Catalogue H61 field. The AB pair is not resolved by the HST either
(Ghez et al. \cite{Ghez et al. 1997a}). There is no evidence in the Hipparcos
astrometric intermediate data that a double solution can be obtained, and
the published solution cannot be improved.

\begin{table}[h]
\caption{The triple system V410 Tau \label{V410Tab}}
\begin{tabular}{lll}
\hline\hline
V410\ Tau & p.a. ($^{\circ }$) & $\rho $ ('') \\ \hline
B & 182 & 0.074 \\
C & 132 & 0.287 \\ \hline
\end{tabular}
\end{table}


\subsubsection{\object{BP Tau}}


This is one of the few objects in Table \ref{Table1} that Hipparcos did not
flag as a suspected binary star. That it is single down to 0.01 arcsec is
further confirmed by HST observations
(Bernacca et al. \cite{Bernacca et al. 1995}). Closer
binarity would obviously not be detected by Hipparcos, and we can rule out
variability-induced motions of the photocentre to explain the parallax found
in the standard data reduction. An undetected orbital motion cannot be an
explanation either, although a one-year orbital period would obviously
result in a confusion between the orbital and the parallactic motion. We
checked that this assumption would imply unreasonably large~masses (on the
order of 15M$_{\odot}$) for the two components.

Because the star is apparently single, we cannot dismiss BP Tau's apparent
parallax easily. We do not believe, however, that it should be taken at face
value for the following reason. The solution published in the Hipparcos
Catalogue represents the best astrometric fit with normalized $\chi^2$
equal to 1.1. If we now compute a solution where we constrain the parallax
to be that of the Taurus stars, we get a fit with normalized $\chi^2$
equal to 1.2. In other words, the solution is only marginally worse than the
published solution, and a true parallax at $2\sigma$ from the published
parallax is not unlikely. Also, one should note that the star, although it
would be located in front of the Taurus group if the Hipparcos parallax were
correct, has the same proper motion as confirmed members of the Taurus SFR.
This casts additional doubt on the published parallax value.

Assuming for the moment that the Hipparcos parallax is correct, is it
plausible that the current location of BP Tau could be explained by its
relatively large heliocentric radial velocity, $14.0\pm 3.0$~km~s$^{-1}$
(Barbier-Brossat \& Figon 1999)? This velocity translates to a LSR velocity
of about +5~km~s$^{-1}$. 
Given the LSR radial velocity of the local molecular cloud
of +7.1~km~s$^{-1}$ (Herbig \cite{Herbig 1977}), 
the radial velocity of BP Tau with respect to the cloud is 
about -2~km~s$^{-1}$. The radial displacement of the star
in its estimated lifetime of 1 Myr (Siess et al. \cite{Siess et al 1999})
is thus about 2
pc. This is obviously not consistent with location of BP Tau in the
molecular cloud at birth.

There is no obvious reason to dismiss the Hipparcos parallax for this star,
but conversely, there is no strong reason to believe it either; the large
statistical error on the result precludes a firm conclusion. The reason for
this large error is not obvious either. Most likely, the culprits are the
faintness of BP Tau and its intrinsic variability. As in the case of DF
Tau, the conclusion is somewhat frustrating, as no clear-cut answer can be
given to the question of these stars' distance. A major conclusion that was
drawn, however, is worth being repeated here: whereas distances of
individual TTSs must be viewed with caution, the distance of Taurus TTSs
\textit{as a group} is entirely consistent with the molecular cloud distance.


\subsubsection{\object{GW Ori}}


One of the brightest CTTSs, GW\ Ori was found to be a single-lined
spectroscopic binary by Mathieu et al. (\cite{Mathieu et al. 1991}), who
give its orbital
parameters. Because of the large distance of this star, the astrometric
perturbation due to orbital motion (reflex motion $\approx 0.5$~mas) is too
small for Hipparcos to detect it.


\subsubsection{\object{AK Sco}}


This is a SB2 system with well-defined orbital parameters (cf. 
Andersen et al. \cite{Andersen et al 1989}).
Unfortunately, the two components have nearly equal
masses and luminosities, so that the photocentre orbital motion is not
significant, precluding a computation of the other orbital elements from
Hipparcos data.


\subsection{Conclusion}

Astrometry turned out to be a powerful tool for detecting the binarity of
variable stars using the variability-induced motion of the photocentre
(Wielen \cite{Wielen 1996}). In spite of the limitations of the method,
which must assume
that only one component is variable and neglects all other causes of
photocentre motions, the binary parameters derived for VIM systems are in
remarkable agreement with other observations. VIM detection of the RY\ Tau
binarity is a long awaited, but somewhat unexpected result of the Hipparcos
mission.

%
\begin{acknowledgements}
%
This research made heavy use of the SIMBAD database and VizieR service of
the Centre de Donn\'{e}es astronomique de Strasbourg (CDS) and of NASA's
Astrophysics Data System Abstract Service. We acknowledge useful discussions
with Sylvie Cabrit, Suzan Edwards and Catherine Turon, 
and thank Jean-Louis Falin for his reanalysis of
\object{CV Cha} and \object{CW Cha} Hipparcos data. We are also indebted to
Joli Adams for smoothing out the English.
%
\end{acknowledgements}
%


\end{document}